# Radiofrequency disorientation in migratory birds: radical pairs or a separate sensory system?

P. J. Hore
Department of Chemistry, University of Oxford, Oxford, UK

peter.hore@chem.ox.ac.uk

ORCID: 0000-0002-8863-570X

## Abstract

Alongside other directional cues, migratory songbirds use the Earth's magnetic field to help them navigate the hundreds or thousands of kilometres between their breeding and over-wintering sites. If it seems surprising that an animal can extract information from a magnetic field a hundred times weaker than a small refrigerator magnet, it should be astounding that this ability can be disrupted by radiofrequency magnetic fields thousands of times weaker still. Magnetic fields with strengths between 1 and 100 nT and frequencies between 0.1 and 100 MHz have been found to prevent captive night-migratory songbirds from orienting in the geomagnetic field (25–65 μT, depending on geographic location). Remarkably, the frequency dependence of this disruption agrees closely with predictions for radical pairs in cryptochrome (the leading hypothesis for the mechanism of the avian magnetic compass), whereas the extraordinarily weak fields that cause disorientation are difficult to reconcile with conventional radical-pair theory. This perspective examines the evidence for radiofrequency-induced disorientation and its relationship to the cryptochrome hypothesis. It concludes by considering a suggestion that birds may possess a distinct radiofrequency-sensitive system based on electromagnetic induction.

## Radical pair magnetoreception

Although the possibility that birds might sense the Earth's magnetic field had been considered since the 19th century [1, 2], convincing evidence for a magnetic compass in migratory songbirds emerged only in the 1960s [3, 4]. The potential value of such an ever-present directional cue to a long-distance migrant is readily apparent: northern wheatears, for example, weighing only 25 g, migrate the length of Asia from Alaska to East Africa and back, completing a round trip of some 29,000 km [5]. Yet, despite more than six decades of research, the biophysical mechanism underlying this remarkable sensory ability remains largely unresolved [6-12]. The leading hypothesis, outlined below, proposes that magnetic-field detection originates in magnetically sensitive biochemical reactions.

The basic idea, summarized in Figure 1(a), is that reactant molecules are converted into signalling-state molecules via short-lived intermediates that can follow one of two competing pathways: either returning to the reactant state or proceeding to a signalling state. For such a reaction to form the basis of a magnetic compass sensor, the branching ratio between these pathways—and hence the ultimate yield of the signalling state—must depend on the direction of the magnetic field in which the reaction occurs. Equipped with such a sensor, a bird moving around in the Earth's magnetic field could obtain directional information by registering variations in the signalling-state population.

On the face of it, this seems a ludicrous suggestion because the energy of the interaction of a molecule with the geomagnetic field is millions of times smaller than the strength of a chemical bond. One might think, therefore, that there would be no way that such a feeble interaction could significantly influence the rates at which chemical bonds are broken or formed. However, in the 1970's it became clear that there exists a class of short-lived, magnetically sensitive reaction intermediates, known as radical pairs, for which this thermodynamic argument is irrelevant. Over the last 50 years, there have been hundreds of *in vitro* studies of the effects of applied magnetic fields on radical-pair reactions [13-17] – effects that can be qualitatively and often quantitatively understood in terms of the quantum mechanics of spin angular momentum.

Radicals are molecules that contain an unpaired electron whose spin is described by a quantum number $S$ with value ½. The total spin of a radical pair depends on whether the spins of the two electrons, one in each radical, are antiparallel ($\uparrow\downarrow$, $S = 0$) or parallel ($\uparrow\uparrow$, $S = 1$). These two forms are known as singlet and triplet, respectively. When the radical pair is created, for example by the transfer of an electron from one molecule to another, it oscillates coherently back and forth between its singlet and triplet states, under the influence of local magnetic fields produced by the nuclei of nearby atoms such as hydrogen and nitrogen (hyperfine interactions; Figure 1(b)). The oscillations, known as quantum beats, typically occur with a range of frequencies up to ~100 MHz, (Figure 1(c)) determined by the quantized energy levels of the electron-nuclear spin system (Figure 1(d)).

The quantum beats are also sensitive to an external magnetic field through the Zeeman interaction with the magnetic moments of the electrons. Consequently, the time-dependent probabilities of the radical pair occupying the singlet and triplet states, and hence the ultimate yield of the signalling state, encode the strength and direction of the external magnetic field.

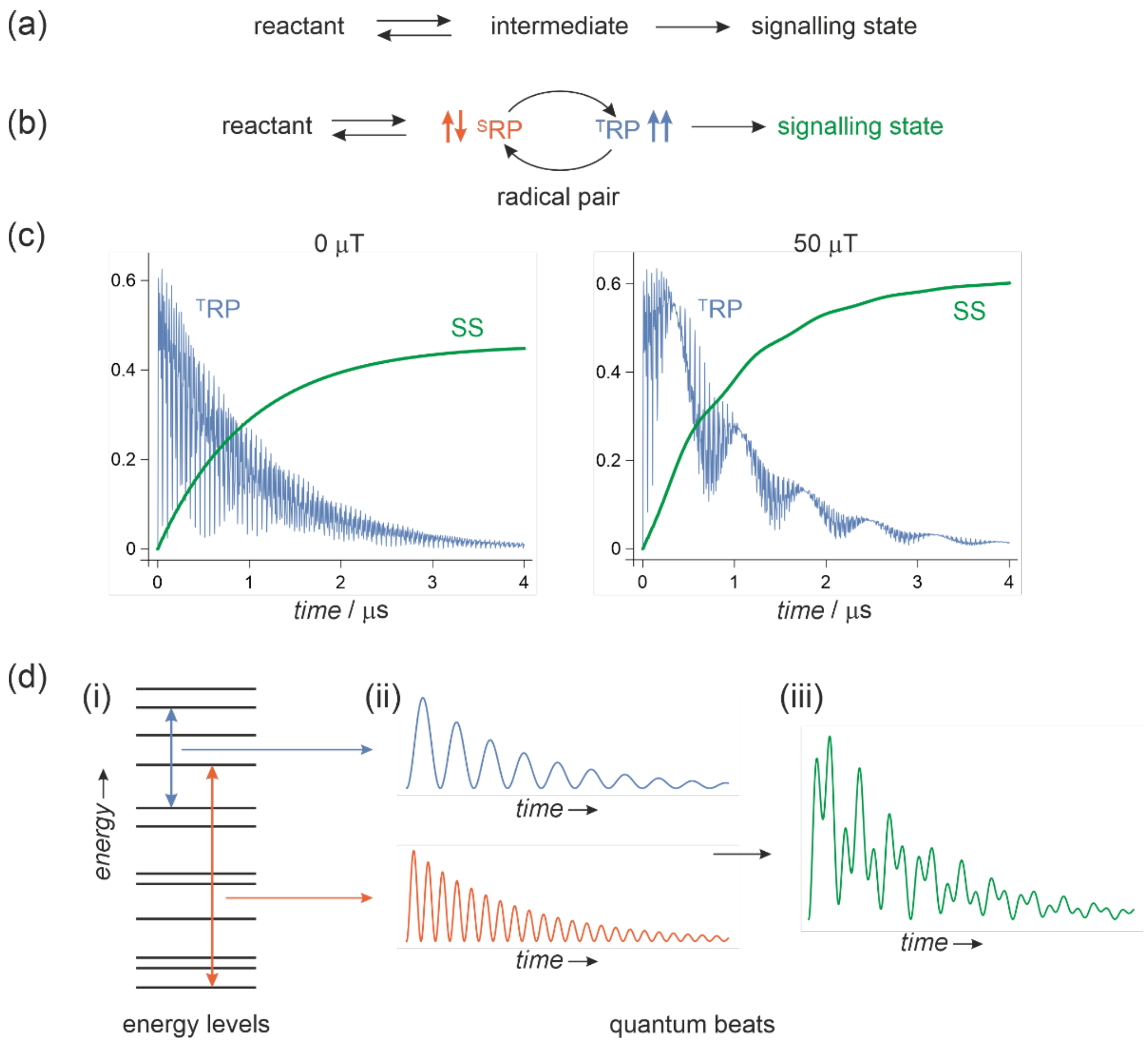

**Figure 1**. **Radical-pair magnetoreception**.

(a) Generic reaction scheme for a chemical compass.

(b) Chemical reaction with a radical-pair (RP) intermediate showing the oscillatory interconversion of singlet ($^{S}$RP) and triplet ($^{T}$RP) states and their spin-selective transformations.

(c) Spin dynamics calculated for a simple model of a radical pair reaction showing the quantum beats in the time-dependence of the $^{T}$RP state (blue) and the build-up of the signalling state (green), without (left) and with (right) an Earth-strength (50 μT) magnetic field. In this case, the magnetic field, on average, increases the fraction of radical pairs in the $^{T}$RP state and hence, also, the final amount of the signalling state.

(d) (i) Schematic spin energy-levels of a radical pair, with arrows indicating two representative coherences. For an energy-level separation $\Delta E$, the corresponding oscillation frequency is $\Delta E / h$ where $h$ is Planck's constant. (ii) These coherences give rise to oscillations in the triplet-state population which are damped by spin relaxation and radical-pair reactions. (iii) Superposition of the individual oscillatory components produces quantum beats of the form illustrated in (c).

In a prescient article in 1978, Schulten proposed a radical-pair basis for the avian magnetic compass [18], compatible with an earlier finding that birds sense the inclination, rather than the polarity, of the geomagnetic field [19]. Schulten also suggested that light could provide the energy necessary to generate radical pairs, implying a light-dependent compass sense, a prediction subsequently confirmed in behavioural experiments [20, 21]. More recently, it has been shown that the magnetoreceptors are located in the birds' retinas [22-24], probably in the outer segments of the double-cone photoreceptor cells [25-27], and that they send information to a brain area that processes visual information [22-24].

Further details of the radical pair magnetoreception hypothesis can be found in several review articles [6-12], particularly in [6] which provides a tutorial explaining "the chemical and physical aspects of the radical-pair mechanism to biologists and the biological and chemical aspects to physicists". The following section outlines the properties of the molecules thought to host the sensory radical pairs. The intention is not to provide a critical review but to set the scene for a discussion of radiofrequency field effects.

## Cryptochrome hypothesis

Cryptochromes (CRY), a class of photoreceptor proteins identified in plants in the 1990's [28], were first proposed as magnetoreceptors more than 20 years ago [29, 30]. A quarter of a century later, no other molecule has emerged as a compelling candidate. Of the six known cryptochromes in birds, CRY4a seems the most likely to have a magnetic sensing function [25, 31-36]. It binds a fully oxidised FAD (flavin adenine dinucleotide) chromophore which, when excited with blue light, generates a radical pair by sequential electron-hopping along a chain of four tryptophan (Trp) residues (Figure 2) [33, 34]. The FAD-Trp radical pairs so-formed are initially singlet and interconvert coherently with the triplet. In purified form, CRY4a from the migratory European robin (*Erithacus rubecula*) exhibits many of the properties expected of a magnetic sensor, including a relatively long-lived radical-pair state whose reactivity can be tuned by external magnetic fields [33]. Related properties have been observed for CRY4a from quail (*Coturnix japonica*) [37], chicken (*Gallus gallus*) [33, 38, 39], and pigeon (*Columba livia*) [33, 38, 40], CRYs from *Drosophila* [41, 42] and *Arabidopsis* [43], and for the closely related bacterial DNA repair enzyme, photolyase [44]. Magnetic sensitivity seems to be a generic property of CRYs that bind FAD (not all do [45-47]).

The competing radical-pair reactions required for magnetic sensitivity (Figure 1(b)) are (i) return to the ground state of the protein and (ii) formation of a stabilised radical-pair state, in which the FAD radical has gained a proton and the Trp radical has lost one [33, 34]. Subsequent reduction of the neutral Trp radical is thought to generate the signalling state, containing the $FADH^{\bullet}$ radical [6]. This state inherits the effect of the magnetic field on the radical pair and could persist for minutes *in vivo*, providing ample time for downstream signal transduction [33].

Further details of the cryptochrome hypothesis and the supporting evidence can be found in recent reviews [6-12, 48].

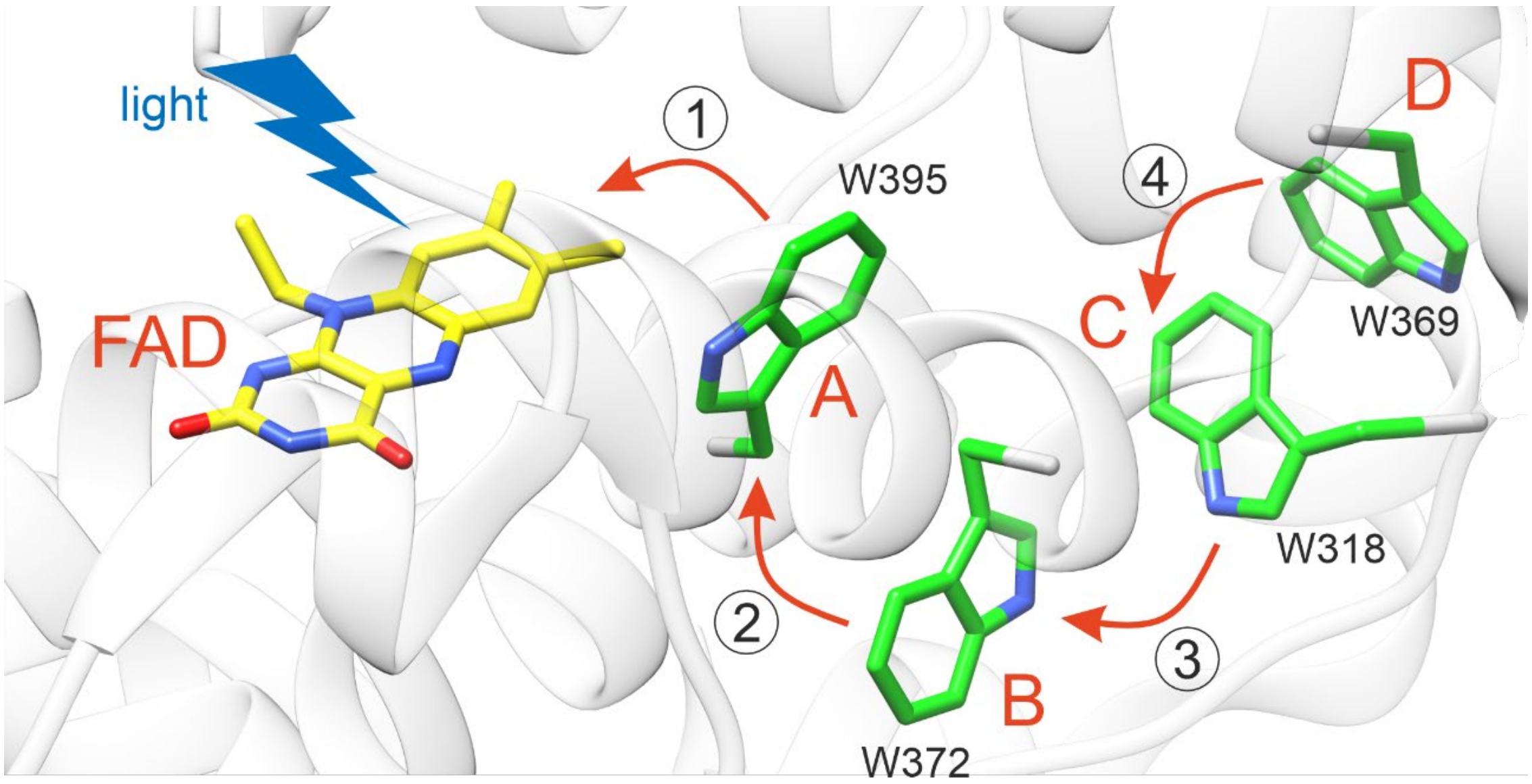

**Figure 2. Cryptochrome photochemistry.**

Structure of the electron transfer chain in pigeon CRY4a [49] showing the FAD chromophore and the tryptophan tetrad ($Trp_A$, $Trp_B$, $Trp_C$, $Trp_D$). Following blue-light excitation of the flavin, the magnetically sensitive FAD-$Trp_D$ radical pair is formed by four sequential electron transfers (red arrows). Adapted from [34].

## Animal detected magnetic resonance

The first evidence that radiofrequency (RF) fields can disrupt the magnetic compass of a migratory songbird emerged in 2004 [50]. When tested in orientation cages known as Emlen funnels [51], European robins were unable to orient in the Earth's magnetic field when exposed either to a single-frequency 7-MHz RF field or to broadband RF spanning 100 kHz–10 MHz. In both cases, the RF field was at least two orders of magnitude weaker than the geomagnetic field.

This behavioural experiment can be seen as an *in vivo* form of "reaction yield detected magnetic resonance" (RYDMR), a well-established technique for studying radical pair reactions *in vitro* [52-58]. As the name suggests, a time-dependent magnetic field stimulates magnetic resonance transitions amongst the energy-levels of the radical pair, so changing its response to a static magnetic field. The difference between the robin experiment and conventional RYDMR is that the readout is the animal's behaviour (i.e. disorientation) rather than a direct measurement of the change in the reaction yield. The robin experiment could therefore be termed "animal-detected magnetic resonance" (ADMR) [58]. The premise underlying ADMR is that the RF field perturbs the spin dynamics of the radical pair sufficiently to degrade the directional information encoded in the signalling state, thereby preventing the birds from orienting using their magnetic compass.

Unlike some other aspects of magnetoreception research, ADMR has withstood the gold-standard test of independent replication. RF disorientation has been observed by three independent research groups (based in Frankfurt, Oldenburg, and St Petersburg), using four species of migratory songbirds: European robin [50, 59-64], garden warbler (*Sylvia borin*) [65-68], Eurasian blackcap (*Sylvia atricapilla*) [69-72], and pied flycatcher (*Ficedula hypoleuca*) [73, 74]. In Emlen-funnel tests, disorientation has been reported for single-frequency RF fields down to ~1 nT and broadband RF fields as weak as ~1 pT/√Hz (~1 nT root-mean-square strength) at frequencies up to ~100 MHz [70, 71, 75].

As a result, ADMR has come to be regarded as a diagnostic test for the operation of a radical-pair-based magnetic compass mechanism, not just in migratory birds but also in: zebra finches [76-78], insects [79, 80], krill [81], rodents [82-84], turtles [85, 86], bats [87], and humans [88, 89]. A variety of assays were employed in these reports, some of which studied behaviours other than orientation. In a few cases, RF exposure was found to cause re-orientation, rather than disorientation, in: snapping turtles [85], wood mice [82], zebra finches [78], rodents, newts, and houseflies [84]. Some of these reports are more convincing than others. Not all have yet been subjected to independent replication and not all have used fully double-blinded protocols and double-wound magnetic-field coils [90].

The next two sections summarize the frequency- and intensity-dependence of ADMR effects on migratory songbirds in the context of the cryptochrome hypothesis.

## ADMR: frequency-dependence

To alter the yield of a radical-pair reaction, the frequency of the oscillating field must coincide with one or more of the quantum-beat frequencies governing singlet–triplet interconversion (Figure 1(c)), i.e. one or more of the energy-level separations of the radical pair (Figure 1(d)). Many of the studies cited above used RF at the Larmor precession frequency of a free electron in the geomagnetic field (28 kHz/μT, i.e. ~1.4 MHz for a ~50 μT field). The electron spins in a radical pair in cryptochrome are in fact far from isolated, being coupled to each other and to more than 20 nuclear spins [91]. A consequence of this network of interacting spins is that resonant effects can be expected to cover a range of frequencies between a few hundred kHz and ~100 MHz, with nothing to distinguish ~1.4 MHz from any other frequency within this range [6, 92].

The lower-frequency limit of the disruptive RF effect appears to lie between 100 kHz and 450 kHz [69]. Noise-modulated RF spanning 0.1–100 kHz does not impair the birds' ability to use their magnetic compass [69], whereas 20–450 kHz broadband RF at a comparable field strength induces disorientation [62]. The existence of a low-frequency threshold can be rationalised by assuming that a magnetic field oscillating sufficiently slowly as to remain effectively constant over the lifetime of the spin coherence should have little influence on the spin dynamics.

Consideration of the energy-level structure of the FAD-Trp radical pair predicts an upper-frequency limit beyond which an oscillating magnetic field should have negligible effect [70-72, 92]. Referring to Figure 3, and recalling that $\nu = \Delta E / h$ (Figure 1(d)), the highest RF frequency ($\nu$) capable of resonant interaction with the radical pair is determined by the energy difference between its highest and lowest spin energy levels (red arrow in Figure 3(a)). This energy gap can be calculated reliably from the known hyperfine interactions and corresponds to ~116 MHz for the FAD-Trp pair [71]. Fields at higher frequencies (green arrows) should therefore be off-resonance and have no effect, whereas lower-frequency fields (blue arrows) can resonate with one or more of the hundreds of thousands of energy-level separations of the radical pair. Simulations of the energy-level spectrum consequently predict a broadly uniform response from ~116 MHz down to the 100–450 kHz low-frequency limit [70, 71, 92].

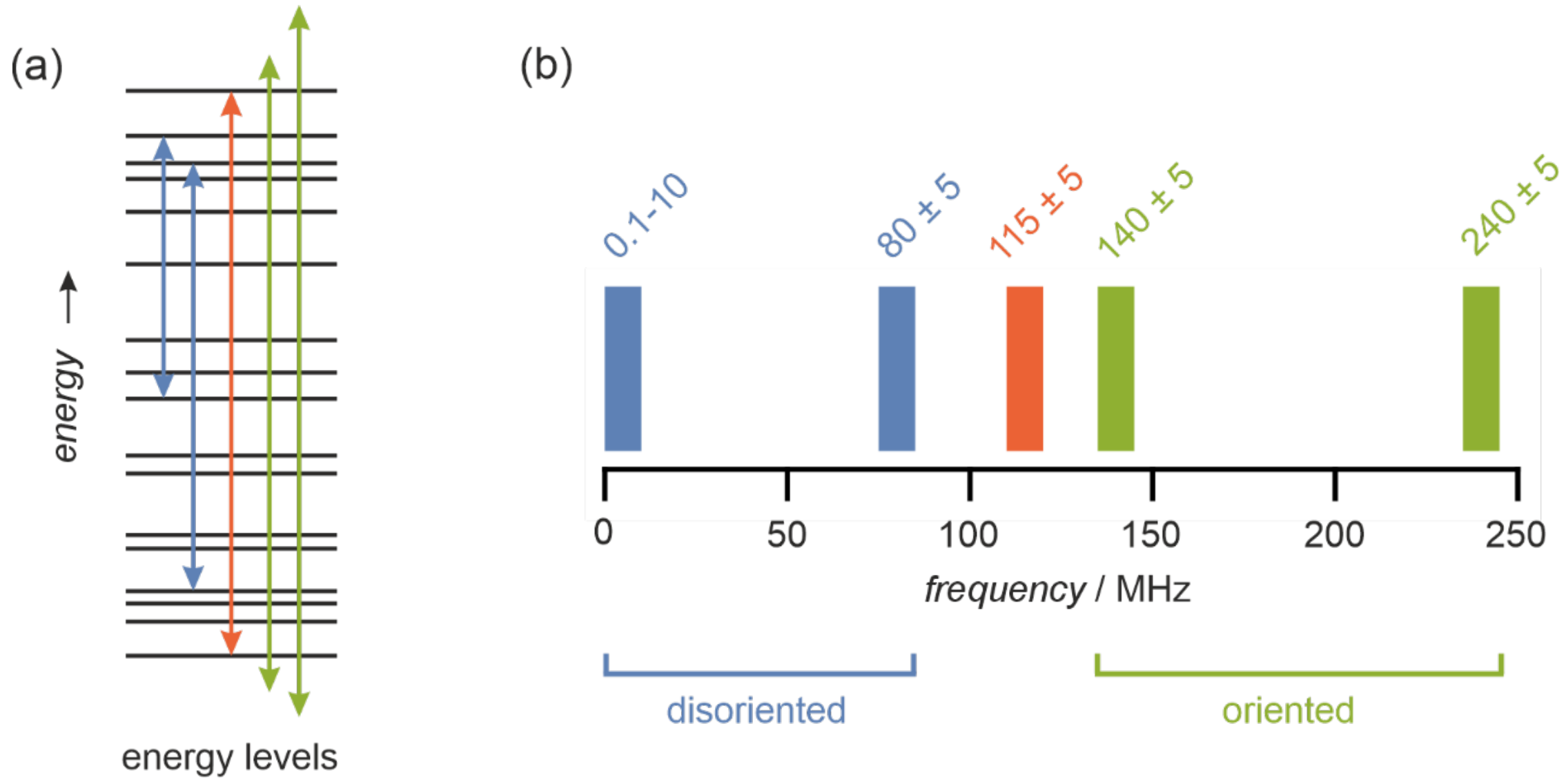

**Figure 3. Radiofrequency disorientation of Eurasian blackcaps.**

(a) Schematic radical-pair energy levels and possible ADMR frequencies.

(b) Results of behavioural tests on Eurasian blackcaps subject to noise-modulated broadband RF fields at different frequencies. Adapted in part from [71].

Figure 3(b) summarizes the results of behavioural experiments on Eurasian blackcaps exposed to 10-MHz-wide, noise-modulated RF fields. Consistent with the quantum-mechanical predictions, fields centred at 5 and 80 MHz caused disorientation [50, 62, 70], whereas fields of comparable intensity centred at 140 and 240 MHz had no detectable effect on the birds' ability to orient [71]. Birds exposed to a frequency band (110–120 MHz) that straddled the predicted cut-off, showed ambiguous behaviour consistent with possible partial disruption [72]. This striking agreement between theory and experiment provides arguably the strongest evidence to date that the avian magnetic compass relies on a FAD-Trp radical pair [71]. More broadly, it represents a rare instance of quantum mechanics successfully predicting an animal's behaviour.

## ADMR: intensity-dependence

To disrupt the directional information provided by a radical-pair sensor, an RF field must not only have an appropriate frequency but must also be sufficiently strong to perturb the populations and coherences of the spin energy levels, thereby altering the extent and timing of singlet–triplet interconversion. To illustrate this point, Figure 4 presents calculations performed for a simplified radical-pair reaction (details in the Supplementary Material). The black line shows the change in signalling-state yield as an Earth-strength, time-independent magnetic field is rotated through 90°, while the red line is the change in this signal induced by a superimposed on-resonance 100-nT RF field. The RF effect is approximately 5000-fold smaller. More realistic models of the FAD-Trp pair predict smaller absolute changes in both geomagnetic and RF field effects, leaving their ratio similar to that in Figure 4. Thus, even at the highest RF field strengths known to disorient migratory songbirds, the predicted perturbation of the signalling-state yield is exceedingly small. For a 1-nT RF field, the expected effect is approximately 50 million times smaller than that of the geomagnetic field. Thus, radical-pair theories that have stood the test of time for 50 years are unable to account for appreciable effects of ≤ 100 nT RF fields.

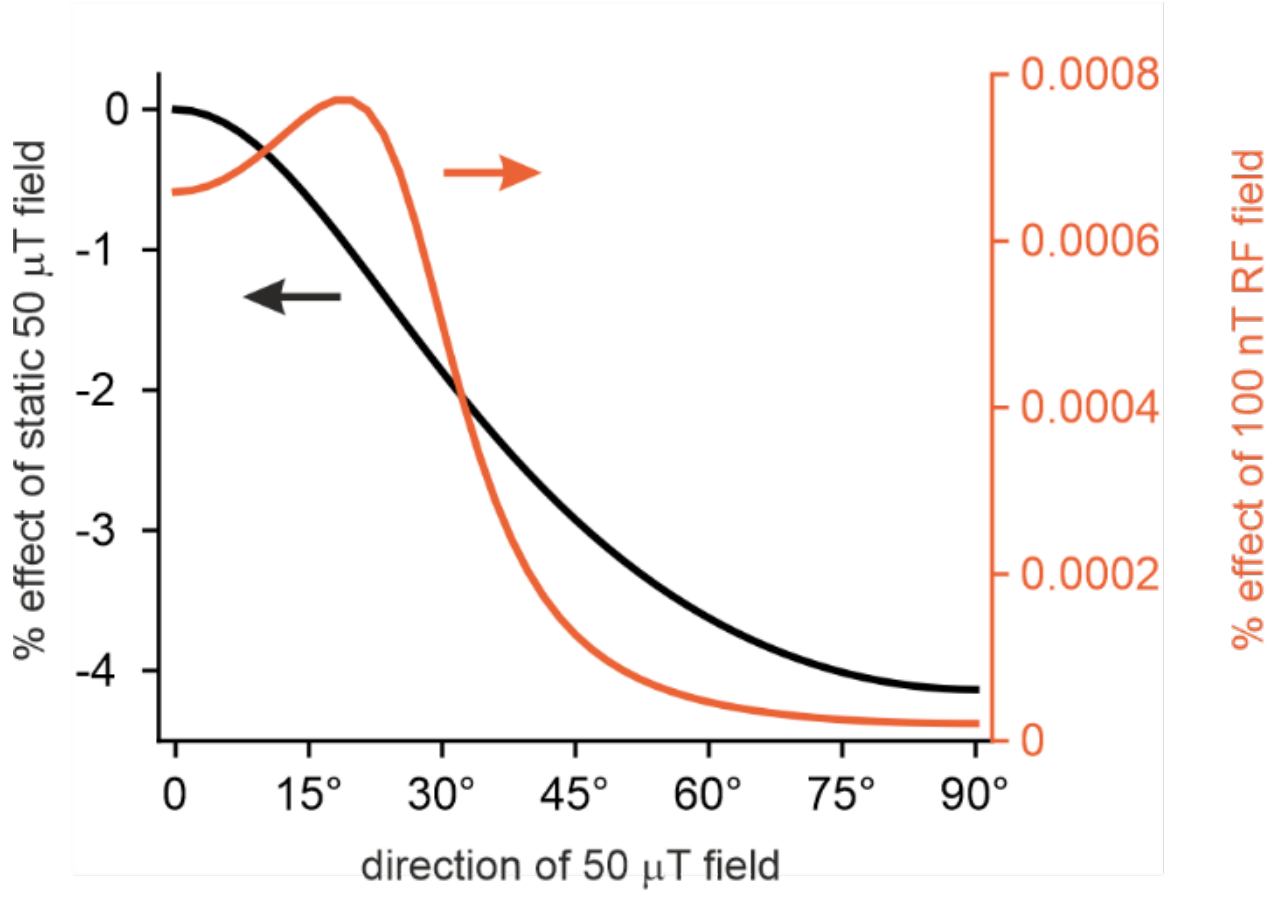

**Figure 4. Calculated radiofrequency magnetic field effect on a simplified radical pair reaction.**

Percentage change in the yield of the signalling state without RF as a function of the direction of a 50 μT static magnetic field (black, left axis) and the percentage change in this signal produced by a resonant 100-nT RF field (red, right axis). The coherence lifetime is 1 μs. See Supplementary Material for details.

Although RF effects have yet to be reported for purified CRY4a, they have been observed in a cryptochrome from a green alga [93] and in engineered flavoproteins derived from the plant photoreceptor, phototropin [93, 94]. However, these experiments employed static magnetic fields more than $10^2$ times stronger than the geomagnetic field and RF fields at least $10^3$ times stronger than the 100-nT upper end of the avian-disorientation range.

## ADMR: discussion

The vast majority of ADMR studies have been predicated on a behavioural response to RF being diagnostic of a radical-pair sensor. As summarized above, this interpretation is strongly supported by the frequency dependence of RF-induced disorientation in blackcaps (Figure 3), yet seems incompatible with the predicted sensitivity of radical-pair reactions to such extraordinarily small perturbations (Figure 4).

Although irreproducible findings are not unknown in magnetoreception research, it seems unlikely that all the reports of RF-induced disorientation of migratory songbirds can be attributed to experimental artefacts. In particular, the independent replication of the basic phenomenon by different research groups using different equipment to conduct the same assay provides confidence that the effect is genuine. This does not, however, establish that RF acts directly on the magnetic sensor: alternative targets could include components of the downstream signal-transduction pathway or mechanisms that influence the birds' motivation to orient in Emlen funnels. However, it would be a remarkable coincidence if an RF-sensitive process mechanistically unrelated to radical pairs exhibited the same frequency response as predicted for the FAD-Trp pair in cryptochrome and observed in behavioural tests on blackcaps (Figure 3).

Generally speaking (see Supplementary Material), an external magnetic field should have an appreciable effect on the spin dynamics of a radical pair when its strength $B$, in microtesla, satisfies $B\tau_{\mathrm{coh}} \geq 20$, where $\tau_{\mathrm{coh}}$ is the spin-coherence lifetime in microseconds. For $\tau_{\mathrm{coh}} \approx 1$ μs (suggested to be optimal for geomagnetic compass sensing [95], $B$ would need to exceed ~20 μT, a value smaller than the geomagnetic field strength but far larger than the 1–100 nT RF fields found to cause disorientation. Conversely, for a 100-nT RF field, $\tau_{\mathrm{coh}}$ would have to be more than ~200 μs. Based on these estimates, there are (at least) three ways in which direct RF interference with a radical pair compass sensor could in theory lead to disorientation.

First, a nanotesla RF field would have time to disrupt the spin dynamics if *the spin coherence persisted for longer than 200 μs*. This seems highly unlikely. Spin decoherence in organic radicals at physiological temperatures is principally driven by modulation of hyperfine interactions by molecular motion [96-98]. Millisecond electron spin relaxation is not uncommon for radicals at cryogenic temperatures where molecular motion is largely frozen but is unheard of at physiological temperatures. Simulations of the molecular and spin dynamics of FAD-Trp radical pairs in cryptochromes suggest that decoherence might be as slow as a microsecond but certainly not hundreds of microseconds [96-98].

Second, assuming the coherence lifetime to be around a microsecond, maybe *the avian nervous system is exquisitely sensitive to tiny fluctuations in the level of the signalling state* such as might be caused by nanotesla RF fields. This seems equally improbable. There is no credible way a 1–100 nT RF field could abolish or render unrecognizable the far stronger directional information derivable from a ~50 μT static field. Given that radical-pair theory struggles to explain quantitatively how a cryptochrome-based light-dependent compass could be sensitive enough to the geomagnetic field [99], it is difficult to imagine how magnetic interactions a thousand times weaker could be significant.

Third, *the current model of radical-pair magnetoreception could be incomplete* in such a way that the spin dynamics of cryptochrome radicals *in vivo* is unprecedentedly and disproportionately vulnerable to feeble time-dependent magnetic interactions. Hypothetically, RF fields might be selectively amplified *in vivo* prior to detection, or spin coherence could be preserved for milliseconds. If such things do occur *in vivo*, there has been no evidence of them from 50 years of *in vitro* research on radical-pair chemistry and physics. Although a number of variants of the conventional radical pair paradigm have been proposed that might boost the expected response to a static magnetic field [99-103], none seem to offer the required thousand-fold boost in RF sensitivity.

In short, there is currently no plausible way to account for a significant effect of nanotesla RF on a geomagnetic radical-pair sensor. Those who study avian magnetoreception, including the author, have responded by suspending disbelief, or ignoring the "elephant in the room", and pressing on with behavioural tests of the cryptochrome hypothesis in the hope that the puzzle will eventually be resolved.

## A separate radio-wave sensor

If radical pairs are not viable detectors of RF fields, an alternative possibility is that migratory birds have evolved a distinct sensory system responsive to environmental RF fields [104], sources of which include thunderstorms (lightning) and solar activity (geomagnetic storms resulting from coronal mass ejections) [75]. Such a system could provide advance warning of adverse terrestrial weather or of geomagnetic disturbances arising from space weather, both of which could be relevant to migration; in the latter case, geomagnetic instability could compromise the reliability of magnetic compass information. A sensor adapted to detect natural RF signals might also respond to anthropogenic RF fields, potentially accounting for the disorientation observed in Emlen-funnel experiments where the only directional cues are magnetic [62, 64].

The concept of a distinct RF sensor was originally raised by Kirschvink who favoured a detection mechanism involving particles of magnetic iron oxide, such as magnetite ($Fe_3O_4$) [104]. The idea was revived in 2026, with electromagnetic induction as a potential explanation [74, 105], building on earlier suggestions that it could underlie geomagnetic sensing [2, 106-109]. Keays and colleagues have proposed that induction could enable the semicircular canals of the avian inner ear, normally associated with vestibular sensing, to function as magnetic field detectors [108, 109]. Supporting evidence includes the discovery of specialised type II hair cells in the semicircular cristae that express highly sensitive voltage-gated ion channels, together with whole-brain mapping studies revealing magnetically induced neuronal activity in brain regions associated with vestibular function [109]. Because the electromotive force generated by induction is proportional to the rate of change of the magnetic field, such a mechanism could, in principle, be far more sensitive to RF fields oscillating at 0.1–100 MHz than to changes in the geomagnetic field produced by head movements at frequencies of the order of 1 Hz [74, 105].

Kavokin et al. [74] suggest that the frequency dependence of RF-induced disorientation could arise from the combined effects of electromagnetic induction, frequency-dependent electrical leakage and rectification by membrane ion channels. Their model predicts a broad sensitivity maximum near 5 MHz, with approximately equal responses at the observed lower and upper frequency limits of 0.4 and 116 MHz. Kattnig [105] likewise argues that Faraday induction could account for the sensitivity to weak RF fields, although the origin of the upper cut-off remains speculative. Both explanations depend on unverified assumptions about electrical coupling and high-frequency signal transduction in the avian inner ear and an unknown biological detection threshold. In short, electromagnetic induction might help explain the RF intensity dependence of the ADMR data, but its ability to explain the frequency dependence is much less firmly established.

None of these considerations argues against the cryptochrome radical-pair hypothesis as the basis of magnetic compass sensing.

## Implications of RF-sensitivity

In a review of the potential effects of 0.1–10 MHz environmental RF noise on magnetoreception, Granger *et al*. [75] concluded that atmospheric, but not solar, RF noise can exceed the levels known to disrupt magnetic orientation in laboratory studies. Across the 1–300 MHz range relevant here (Figure 3), natural RF noise generated by lightning generally decreases strongly with increasing frequency. Although the ambient background is dominated by lower-frequency emissions, rapid electrical breakdown processes can generate intense, short-duration bursts extending well into the VHF band (30–300 MHz) [110]. Solar radio emission exhibits a more complex frequency dependence: quiet-Sun spectral flux density generally increases with frequency in the VHF band, whereas solar radio bursts can produce much stronger, highly variable emissions spanning a broad spectrum [111, 112]. At ground level, solar emissions below approximately 10 MHz are largely blocked by the terrestrial ionosphere, while higher-frequency emissions can contribute to the natural RF background, particularly during periods of enhanced solar activity [113]. These observations suggest that naturally occurring RF noise associated with terrestrial and space weather phenomena may reach intensities sufficient to disrupt magnetic orientation in birds.

The RF-induced disorientation observed in behavioural experiments raises important questions about its potential ecological consequences for migratory birds. European robins tested in unscreened wooden huts on the campus of the University of Oldenburg were unable to orient using their magnetic compass in the presence of ambient anthropogenic electromagnetic noise, often referred to as electrosmog [62]. When the huts were lined with grounded aluminium sheets, attenuating the electromagnetic noise by approximately two orders of magnitude, the birds were able to orient appropriately. Removing the grounding, or introducing artificial RF noise into the shielded huts, again disrupted their orientation. By contrast, birds tested at a rural site away from substantial electrosmog sources oriented normally without shielding [62].

Whether natural or anthropogenic RF noise has measurable effects on the migration routes, headings, navigational success, or survival of free-flying birds, however, remains unclear [114]. Under natural conditions, birds have access to multiple sources of directional information, including celestial cues, and may therefore assign little or no weight to magnetic compass information when it is judged to be unreliable or disrupted.

RF interference should be considered a potential confounding factor in laboratory studies of animal magnetoreception. Commonplace electrical equipment, wiring and grounding can introduce unintended RF noise [84, 105]. Ambient RF levels should therefore be carefully measured and reported, and minimized through appropriate shielding, filtering and grounding, with comparable RF environments maintained during holding/training and testing [75, 84, 92, 105]. Such precautions may help reduce experimental variability and improve the reproducibility of behavioural tests of magnetoreception.

## Outlook

One way to distinguish whether RF sensitivity in migratory songbirds arises from a direct effect on the radical-pair compass or from a separate RF-detector would be to alter the magnetic properties of CRY4a in the retina, thereby shifting the frequencies predicted to cause disorientation. A corresponding shift in behavioural responses to RF would strongly support a direct effect on the radical-pair sensor; an unchanged response would favour a separate radio-wave detector. Such a test might be feasible using isotopic substitution to alter the hyperfine interactions that govern external magnetic field effects [115, 116].

A recent computational study of radical-pair magnetoreception predicted that a single-frequency RF field, acting alone, could provide directional information comparable to that supplied by a static magnetic field of similar strength, potentially allowing birds to orient in an oscillating field in the absence of the geomagnetic field [57]. If confirmed experimentally, this prediction would provide evidence against an induction-based RF sensor, which is unlikely to have evolved to detect the direction of oscillating magnetic fields.

A third approach would be to determine whether RF fields act directly on the magnetic-sensing pathway in the retina and whether any such effect depends on CRY4a. This will require detection of magnetically induced changes in retinal neuronal activity, for example by electrophysiology [117-120] or calcium imaging [121, 122]. If successful, such recordings could be extended to include RF stimuli [58]. Confirming a specific role for CRY4a would additionally require altering its expression or function in the retina. Although such manipulations in migratory songbirds remain challenging [37], the recent development of adeno-associated virus-mediated gene delivery to European robin photoreceptors provides a promising route towards targeted *in vivo* interventions [123]. Alternatively, a more genetically tractable model organism could be used. Extensive behavioural studies have cast doubt on the magnetic sensitivity of *Drosophila* [124-127], but the Bogong moth (*Agrotis infusa*), a migratory insect with a magnetic compass, may be more promising, particularly as behavioural and electrophysiological methods for studying its compass systems are already well developed [80, 128, 129].

The extraordinary distances travelled by migratory birds expose them to a wider range of threats than those faced by many resident species. Protecting birds that cross multiple national boundaries is particularly challenging, especially in the face of pressures such as habitat destruction and climate change. Translocating migrants away from degraded habitats is rarely successful, as their powerful navigational instincts often drive them back to their original capture sites. A deeper mechanistic understanding of how these remarkable navigators find their way could open up new possibilities for conservation, potentially allowing researchers to ‘trick’ migratory birds into accepting safer locations as their new homes.

## Acknowledgements

I am pleased to acknowledge generous funding from the European Research Council (Synergy Grant, *QuantumBirds*, grant agreement 810002, under the European Union’s Horizon 2020 research and innovation programme) and the Deutsche Forschungsgemeinschaft (Magnetoreception and navigation in vertebrates, SFB 1372, project no. 395940726). I am deeply grateful to the many collaborators, postdocs and students with whom I have had the privilege of working over the past 50 years.

# Supplementary Material

## Radiofrequency disorientation in migratory birds: radical pairs or a separate sensory system?

P. J. Hore
Department of Chemistry, University of Oxford, Oxford, UK

peter.hore@chem.ox.ac.uk

## Relation between magnetic-field strength and coherence lifetime

To have a significant effect on the yield of the signalling state, an external magnetic field must alter the spin dynamics of the radical pair before too much of the spin coherence has disappeared. Two routes for the loss of coherence can be identified: spin relaxation, which drives the spin system towards thermal equilibrium (decoherence), and chemical reactions that transform radical pairs into reactants or products. In general, the spin-coherence lifetime, $\tau_{\text{coh}}$, is determined by the faster of these two processes (specifically, the reciprocal of its rate constant).

The characteristic frequency of a free electron in an external magnetic field of strength $B$ is $\gamma_{\text{e}}B$, where the gyromagnetic ratio, $\gamma_{\text{e}} = 1.76\times10^{11}\ \text{T}^{-1}\ \text{s}^{-1}$. For the field to have a significant effect on the spin dynamics, the angle through which the electron spin "rotates" during the lifetime of the spin coherence, $\gamma_{\text{e}}B\tau_{\text{coh}}$, should be greater than $\sim\pi$, i.e. $B\tau_{\text{coh}} \geq \pi/\gamma_{\text{e}} \approx 2\times10^{-11}\ \text{T s}$ or $B\tau_{\text{coh}} \geq 20$, when $B$ is expressed in $\mu$T and $\tau_{\text{coh}}$ in $\mu$s.

## Figure 4

The simulations shown in Figure 4 were performed using the γ-COMPUTE method [1-3] with the following parameters.

The quantity calculated is the fractional yield of the product formed from the triplet radical pair state, $\Phi_{\text{T}}(\theta, B_0, B_1)$, averaged over the initial RF phase, $0 \leq \gamma \leq 2\pi$.

*Spin system*: 2 electron spins, each coupled to one spin-1/2 nuclear spin.

*Hyperfine interactions*: $\mathbf{A}_1 = \begin{pmatrix} 0 & 0 & 0 \\ 0 & 0 & 0 \\ 0 & 0 & 1.0 \end{pmatrix} \text{mT}$; $\mathbf{A}_2 = \begin{pmatrix} 0 & 0 & 0 \\ 0 & 0 & 0 \\ 0 & 0 & 0.3 \end{pmatrix} \text{mT}$.

*Dipolar interaction*: $\mathbf{D} = D\begin{pmatrix} -2/3 & 0 & 0 \\ 0 & -2/3 & 0 \\ 0 & 0 & 4/3 \end{pmatrix}$ with $D = -400\ \mu\text{T}$

*Exchange interaction*: zero.

*Static magnetic field*: $\mathbf{B}_0 = B_0 \begin{pmatrix} \sin\theta \\ 0 \\ \cos\theta \end{pmatrix}$.

*Radiofrequency field*: $\mathbf{B}_1 = B_1 \sin\left(2\pi \nu_{\mathrm{RF}} t + \gamma\right) \begin{pmatrix} 1 \\ 0 \\ 0 \end{pmatrix}$.

*Radiofrequency*: $\nu_{\mathrm{RF}}$ = 9.05 MHz (the frequency at which a $B_1$ = 100-nT RF field produces the greatest change in $\Phi_{\mathrm{T}}(0,B_0,B_1)$, with $B_0$ = 50 μT.

*Recombination rate constants*: $k_{\mathrm{S}} = k_{\mathrm{T}} = k = 10^6\ \mathrm{s}^{-1}$ (symmetric reactions of singlet and triplet pairs to give distinct products). Because spin relaxation was not included, the spin-coherence lifetime is simply the radical-pair lifetime, i.e. $\tau_{\mathrm{coh}} = k^{-1} = 1\ \mu\mathrm{s}$.

γ-COMPUTE *nodes*: $n = 128$.

The curves in Figure 4 are:

Black: $100\% \times \dfrac{\Phi_{\mathrm{T}}(\theta,B_0,0) - \Phi_{\mathrm{T}}(0,0,0)}{\Phi_{\mathrm{T}}(0,0,0)}$ with $B_0 = 50\ \mu\mathrm{T}$ and $0 \le \theta \le 90°$.

Red: $100\% \times \dfrac{\Phi_{\mathrm{T}}(\theta,B_0,B_1) - \Phi_{\mathrm{T}}(\theta,B_0,0)}{\Phi_{\mathrm{T}}(0,0,0)}$ with $B_0 = 50\ \mu\mathrm{T}$, $B_1 = 100\ \mathrm{nT}$, and $0 \le \theta \le 90°$.

## Supplementary References